\pdfoutput=1
\documentclass[usenatbib]{mn2e}
\usepackage{apjfonts}
\usepackage{amssymb}
\usepackage{amsmath}
\usepackage{ctable}
\usepackage{url}
\usepackage{fixltx2e} 

\newcommand{\be}{\begin{equation}}
\newcommand{\ee}{\end{equation}}

\newcommand{\msun}{M_{\sun}}

\newcommand{\movieurltwo}{\url{http://www.cfa.harvard.edu/~phopkins/Site/Movies_zoom.html}}

\newcommand\plotone[1]
 {\centering \leavevmode \includegraphics[width={0.99\columnwidth}]{#1}}

\newcommand{\acknowledgments}{\begin{small}\section*{Acknowledgments}\end{small}}
\newcommand\altaffilmark[1]{$^{#1}$}
\newcommand\altaffiltext[1]{$^{#1}$}
\title[The Origin of the Eccentric Disk and BH in M31]{The Nuclear Stellar Disk in Andromeda: A Fossil from the Era of Black Hole Growth}

\author[Hopkins and Quataert]{
\parbox[t]{\textwidth}{ 
Philip F. Hopkins\altaffilmark{1}\thanks{E-mail:phopkins@astro.berkeley.edu} \&
Eliot Quataert\altaffilmark{1}} 
\vspace*{6pt} \\
\altaffiltext{1}{Department of Astronomy and Theoretical Astrophysics
  Center, University of California
  Berkeley, Berkeley, CA 94720} \\
}

\date{Submitted to MNRAS, February 2, 2009}
\begin{document}
\maketitle
\label{firstpage}
\begin{abstract}
  The physics of angular momentum transport from galactic scales
  ($\sim 10-100$ pc) to much smaller radii is one of the oustanding
  problems in our understanding of the formation and evolution of
  super-massive black holes (BHs). Seemingly unrelated observations
  have discovered that there is a lopsided stellar disk of unknown
  origin orbiting the BH in M31, and possibly many other systems.  We
  show that these nominally independent puzzles are in fact closely
  related.  Multi-scale simulations of gas inflow from galactic to BH
  scales show that when sufficient gas is driven towards a BH,
  gravitational instabilities form a lopsided, eccentric disk that
  propagates inwards from larger radii.  The lopsided stellar disk
  exerts a strong torque on the remaining gas, driving inflows that
  fuel the growth of the BH and produce quasar-level luminosities. The
  same disk can produce significant obscuration along many sightlines
  and thus may be the putative ``torus'' invoked to explain obscured
  active galactic nuclei and the cosmic X-ray background.  The stellar
  relic of this disk is long lived and retains the eccentric
  pattern. Simulations that yield quasar-level accretion rates produce
  relic stellar disks with kinematics, eccentric patterns, precession
  rates, and surface density profiles in reasonable agreement with
  observations of M31.  The observed properties of nuclear stellar
  disks can thus be used to constrain the formation history of
  super-massive BHs.
\end{abstract}

\begin{keywords}
  galaxies: active --- quasars: general --- galaxies: evolution ---
  cosmology: theory
\end{keywords}

\section{Introduction}
\label{sec:intro}

A massive black hole (BH) resides at the center of most massive
galaxies
\citep{KormendyRichstone95,Gebhardt00,merrittferrarese:msigma}.  Such
BHs gain most of their mass as luminous quasars
\citep{soltan82,yutremaine:bhmf}.  A long-standing problem in
understanding the origin of massive BHs is how gas loses angular
momentum and inflows from galactic scales all the way to the BH.
Moreover, the self-gravity of gas can cause it to locally collapse and
turn into stars; whatever process drives inflow must compete against
gas consumption via star formation.  It is now well-established that
on relatively large scales within a galaxy, disturbances from
collisions with other galaxies and global self-gravitating
instabilities can bring the gas down to radii ($\sim10-100\,$pc) where
the direct gravitational force of the BH begins to dominate the
dynamics \citep{shlosman:bars.within.bars,
  schweizer98,barnes:review}. However, the BH also efficiently
suppresses the disturbances from larger scales.  How, then, does the
gas continue to flow in?  On the smallest scales ($\ll 0.1\,$pc),
accretion can occur through angular momentum transport by local
magnetic stresses \citep{balbus.hawley.review.1998}. But this leaves a
critical gap of a factor of $\sim10^{2-3}$ in radius, in which gas is
still weakly self-gravitating and can form stars, but both
larger-scale torques and local magnetic stresses are inefficient;
models have traditionally had great difficulty in crossing this gap
\citep{shlosman:inefficient.viscosities,
  goodman:qso.disk.selfgrav,thompson:rad.pressure}.

Independent observations of the properties of stars close to BHs in
nearby galaxies have, in some cases, discovered that many of the old
stars reside in an eccentric, lopsided stellar disk on spatial scales
from $\sim1-10\,$pc \citep{lauer:ngc4486b,lauer:centers,
  houghton:ngc1399.nuclear.disk,
  thatte:m83.double.nucleus,debattista:vcc128.binary.nucleus}. The
most well-known and well-studied case is in the neighbor to the Milky
Way, the Andromeda galaxy, M31
\citep{lauer93,tremaine:m31.nuclear.disk.model,bender:m31.nuclear.disk.obs}.
The dynamics of this disk have received considerable attention, but
its origin remains poorly understood
\citep{peiris:m31.nuclear.disk.models,
  salow:nuclear.disk.models.2,bender:m31.nuclear.disk.obs}.  Given the
demanding resolution requirements, it is also not clear whether such
features are peculiar or generic.
  
To understand the angular momentum transport required for massive BH
growth, we have recently carried out a series of numerical simulations
of inflow from galactic to BH scales
\citep{hopkins:zoom.sims}.\footnote{\label{foot:url}Movies of these
  simulations are available at \movieurltwo} By re-simulating the
central regions of galaxies, gas flows can be followed from galactic
scales of $\sim100\,$kpc to much smaller radii, with an ultimate
spatial resolution $<0.1\,$pc.  For sufficiently gas-rich disky
systems, gas inflow continues all the way to $\lesssim 0.1$ pc.  Near
the radius of influence of the BH, the systems become unstable to the
formation of lopsided, eccentric gas+stellar disks.  This eccentric
pattern is the dominant mechanism of angular momentum transport at
$\lesssim 10$ pc, and can lead to accretion rates as high as
$\sim10\,\msun\,{\rm yr^{-1}}$, sufficient to fuel the most luminous
quasars.  In addition, through this process, some of the gas
continuously turns into stars and builds up a nuclear stellar disk.
In this {\em Letter}, we examine the possibility that the nuclear
stellar disks seen in M31 and other galaxies are ``fossils'' from the
era of BH growth.  If correct, this provides a powerful new set of
constraints on the formation and evolution of supermassive BHs.

\vspace{-0.7cm}
\section{The Simulations}
\label{sec:sims}

\citet{hopkins:zoom.sims} give a detailed description of the
simulations used here; we briefly summarize some of their most
important properties.  The simulations were performed with the
parallel TreeSPH code {\small GADGET-3} \citep{springel:gadget}.
The simulations include collisionless stellar disks and bulges, dark
matter halos, gas, and BHs.  For this study, we are interested in
isolating the physics of gas inflow. As a result, we do not include
models for BH accretion and feedback -- the BH's mass is constant in
time and its only dynamical role is via its gravitational influence on
scales $\lesssim 10\,$pc.

Because of the very large dynamic range in both space and time needed
for the self-consistent simulation of galactic inflows and nuclear
disk formation, we use a ``zoom-in'' re-simulation approach.  This
begins with a large suite of simulations of galaxy-galaxy mergers, and
isolated bar-(un)stable disks.  These simulations have
$0.5\times10^{6}$ particles, corresponding to a spatial resolution of
$50\,$pc. These simulations have been described in a series of previous papers 
\citep{dimatteo:msigma,robertson:msigma.evolution,
  cox:kinematics,younger:minor.mergers,hopkins:disk.survival}.
From this suite of simulations, we select representative simulations
of gas-rich major mergers of Milky-Way mass galaxies (baryonic mass
$10^{11}\,\msun$), and their isolated but bar-unstable analogues, to
provide the basis for our re-simulations. 
The dynamics on smaller scales does not depend critically on the
details of the larger-scale dynamics.  Rather, the small-scale
dynamics depends primarily on global parameters of the system, such as
the total gas mass channeled to the center relative to the
pre-existing bulge mass.

Following gas down to the BH accretion disk requires much higher
spatial resolution than is present in the galaxy-scale simulations. We
begin by selecting snapshots from the galaxy-scale simulations at key
epochs.
In each, we isolate the central $0.1-1$\,kpc region, which contains
most of the gas that has been driven in from large scales.  Typically
this is about $10^{10}\,\msun$ of gas, concentrated in a roughly
exponential profile with a scale length of $\sim0.3-0.5\,$kpc. 
From this mass distribution, we then re-populate the gas in the
central regions at much higher resolution, and simulate the dynamics
for several local dynamical times.  These simulations involve $10^{6}$
particles, with a resolution of a few pc and particle masses of
$\approx 10^{4}\,\msun$.  We have run $\sim50$ such re-simulations,
corresponding to variations in the global system properties, the model
of star formation and feedback, and the exact time in the larger-scale
dynamics at which the re-simulation occurs.
\citet{hopkins:zoom.sims} present a number of tests of this
re-simulation approach and show that it is reasonably robust for this
problem.  This is largely because, for gas-rich disky systems, the
central $\sim 300$ pc becomes strongly self-gravitating, generating
instabilities that dominate the subsequent dynamics.


These initial re-simulations capture the dynamics down to $\sim 10$
pc, still insufficient to quantitatively describe accretion onto a
central BH.  We thus repeat our re-simulation process once more, using
the central $\sim10-30\,$pc of the first re-simulations to initialize
a new set of even smaller-scale simulations.  These typically have
$\sim10^{6}$ particles,
a spatial resolution of $0.1\,$pc, and a particle mass
$\approx100\,\msun$. We carried out $\sim50$ such simulations to test
the robustness of our conclusions and survey the parameter space of
galaxy properties.  These final re-simulations are evolved for
$\sim10^{7}$ years -- many dynamical times at $0.1$\,pc, but very
short relative to the dynamical times of the larger-scale parent
simulations.\footnote{We also carried out a few extremely
  high-resolution intermediate-scale simulations, which include
  $\sim5\times10^{7}$ particles and resolve structure from $\sim$ kpc
  to $\sim0.3\,$pc -- these are slightly less high-resolution than the
  net effect of our two zoom-ins, but they obviate the need for a
  second zoom-in iteration. The conclusions from these higher
  resolution simulations are identical.}


Our simulations include gas cooling and star formation, with gas
forming stars at a rate motivated by the observed \citet{kennicutt98}
relation. Specifically, we use a star formation rate per unit volume
$\dot \rho_{\ast} \propto \rho^{3/2}$ with the normalization chosen so
that a Milky-way like galaxy has a total star formation rate of about
$1\,M_{\sun} \, {\rm yr^{-1}}$.
Because we cannot resolve the detailed processes of supernovae
explosions, stellar winds, and radiative feedback, feedback from stars
is modeled with an effective equation of state 
\citep{springel:multiphase}. In this model, feedback is assumed to
generate a non-thermal (turbulent, in reality) sound speed that
depends on the local star formation rate, and thus the gas density.
We use sub-grid sound speeds $\sim20-100\,{\rm km\,s^{-1}}$, motivated
by a variety of observations of dense, star forming regions both
locally and at high redshift \citep{downes.solomon:ulirgs,
  bryant.scoville:ulirgs.co,forsterschreiber:z2.disk.turbulence,
  iono:ngc6240.nuclear.gas.huge.turbulence}.
Within this range, we found little difference in the physics of
angular momentum transport or in the resulting accretion rates, gas
masses, etc. \citep{hopkins:zoom.sims} (see also \S \ref{sec:results}).


\vspace{-0.67cm}
\section{Results}
\label{sec:results}

\begin{figure}
    \centering
    \plotone{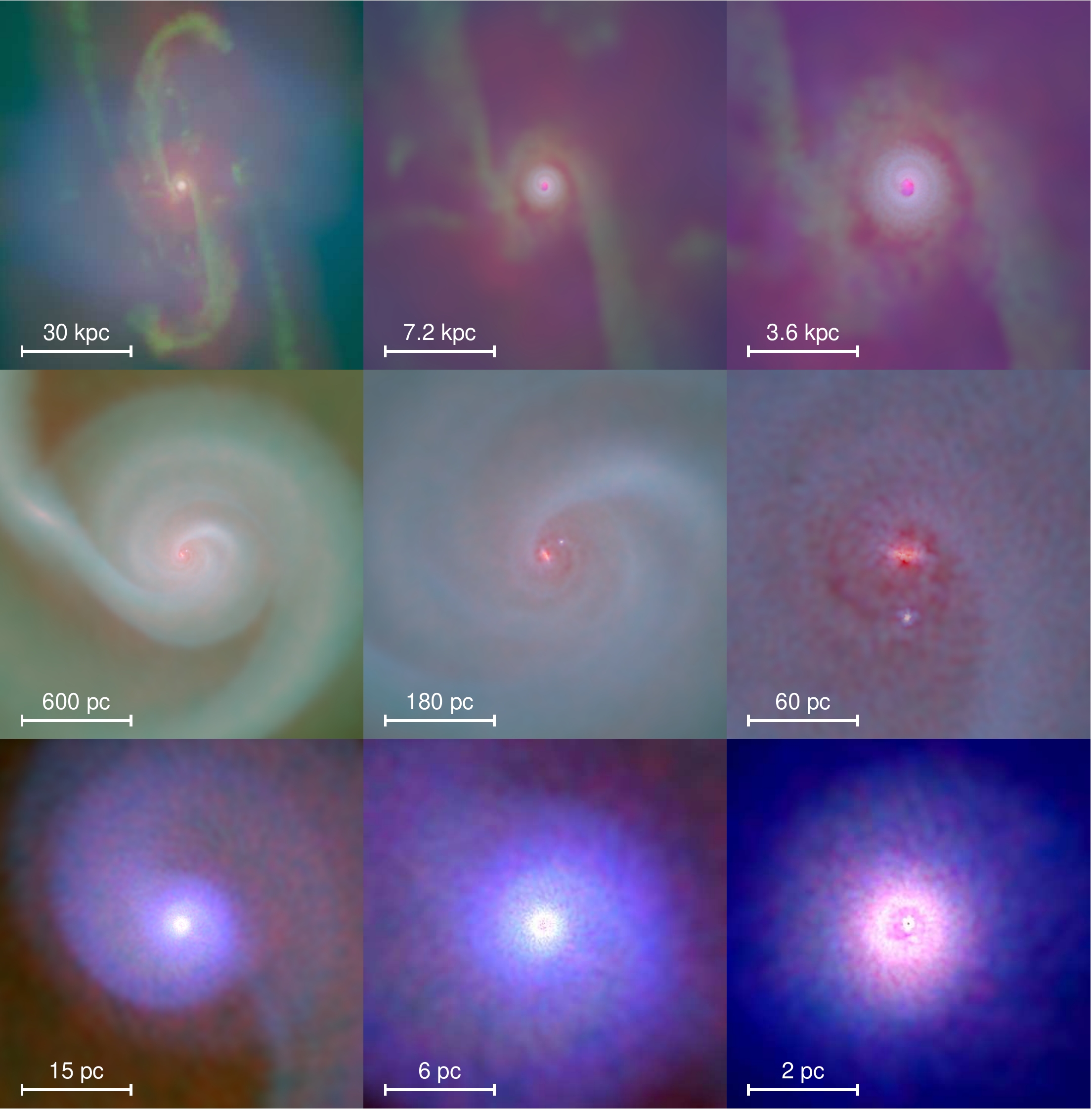}
    \caption{Example of the multi-scale simulations used to follow gas
      inflow from galactic to nuclear scales.  In each panel, red
      colors denote the projected stellar mass density, and green/blue
      colors denote the projected gas density (with the variation from
      green to blue reflecting an increasing star formation rate per
      unit mass in the gas). Each image is rotated to project the gas
      density face on, relative to its net angular momentum vector
      {\em Top:} Galaxy-scale: the merger of two similar-mass
      galaxies, just after the final coalescence of the two nuclei
      (two $10^{11}\,\msun$ galaxies with a disk gas fraction $f_{\rm
        gas}\sim0.4$ at the time of merger, and an initial bulge to
      total mass ratio of $B/T = 0.2$; simulation b3ex(co) in
      \citealt{hopkins:zoom.sims}).  The merger has driven large
      amounts of gas into the central $\sim1\,$kpc, forming the
      nuclear starburst shown.  {\em Middle:} Re-simulation of the
      $\sim 0.1-1$ kpc region (simulation If9b5).  The starburst disk,
      being strongly self-gravitating, develops a spiral and bar mode
      that drives gas to $\sim10\,$pc, where the bar is suppressed by
      the gravity of the BH.  {\em Bottom:} Re-simulation of the
      central $\sim 30$\,pc (simulation Nf8h1c1qs).  The inflow to
      these scales rapidly forms a lopsided eccentric disk around the
      BH (which maps onto a one-armed spiral at larger radii). The
      disk drives accretion rates of $\sim 1-10\,\msun\,{\rm yr^{-1}}$
      to $<0.1\,$pc, and leaves the eccentric stellar relics shown in
      Figures \ref{fig:tile.10pc} \& \ref{fig:velocities}.
      \label{fig:zoom}}
\end{figure}

Figure~\ref{fig:zoom} shows an example of the results of our
re-simulations. We follow the merger and coalescence of two Milky-way
mass galaxies, a highly asymmetric event chosen because it is likely
to lead to rapid BH growth.  The first image (top left) provides a
large-scale view of the system just after the coalescence of the two
galactic nuclei and their central BHs. The highly asymmetric
disturbances visible in the image (e.g.\ tidal tails) efficiently
torque the gas and allow it to flow inwards.  Inside the central
$\sim$kpc, the inflowing gas piles up at the point where its gravity
begins to dominate that of the stars and dark matter; this dense gas
generates a luminous burst of star formation.  Precisely because the
gas and newly-formed stars are self-gravitating, they form secondary
gravitational instabilities such as bars and spiral waves that produce
further torques and inflow. This is essentially the ``bars within
bars'' mechanism proposed in \citet{shlosman:bars.within.bars} and it
occurs for the reasons outlined therein.  But once the gas reaches
$\sim10\,$pc, this mechanism no longer works -- the gravity of the BH
begins to dominate ($M_{BH} = 3\times10^{7}\,\msun$ here) and the
system can no longer support the large-scale bars critical to the
inflow on larger scales.

At precisely these scales our simulations demonstrate that a new
instability generically arises -- a nearly static (slowly precessing)
lopsided or eccentric disk of gas and stars.  Such slowly varying
eccentric ($m=1$) perturbations are unique to the gravitational field
of a point mass such as a BH
\citep{tremaine:slow.keplerian.modes}. But they are also linearly
stable \citep{tremaine:slow.keplerian.modes}, so how do they arise in
the simulations?  We discuss this in detail in
\citet{hopkins:zoom.sims}; to summarize, we believe that the mode is a
{\em global} phenomena that grows from the outside in.  It first
starts to grow because of self-gravity, where the mass of the (stellar
+ gas) disk is comparable to the mass of the BH.  In the simulations,
this occurs at $\sim10-100\,$pc. Gas moving in circular orbits passes
through this eccentric disk, and experiences a torque from the stars
therein, causing the gas to lose angular momentum and fall inwards.
Some of the gas turns into stars, those stars are excited into the
$m=1$ mode, allowing the perturbation to efficiently propagate inwards
to $\sim0.1\,$pc.  Although at radii where $M_{\rm disk}(<R) \ll
M_{\rm BH}$, the system may formally be stable, the eccentric pattern
is {\em induced} by the mass distribution at somewhat larger radii.
The lopsided or eccentric disk, then, is a coherent global pattern
superimposed on the otherwise axisymmetric gas and stellar mass
distribution.  The eccentric pattern precesses with an angular pattern
speed $\Omega_{p}$, which we measure to be $\sim 1-5\,{\rm
  km\,s^{-1}\,pc^{-1}}$ (independent of radius), set by the rotation
rate at the radii where the disk and BH masses are comparable.

\begin{figure}
    \centering
    \plotone{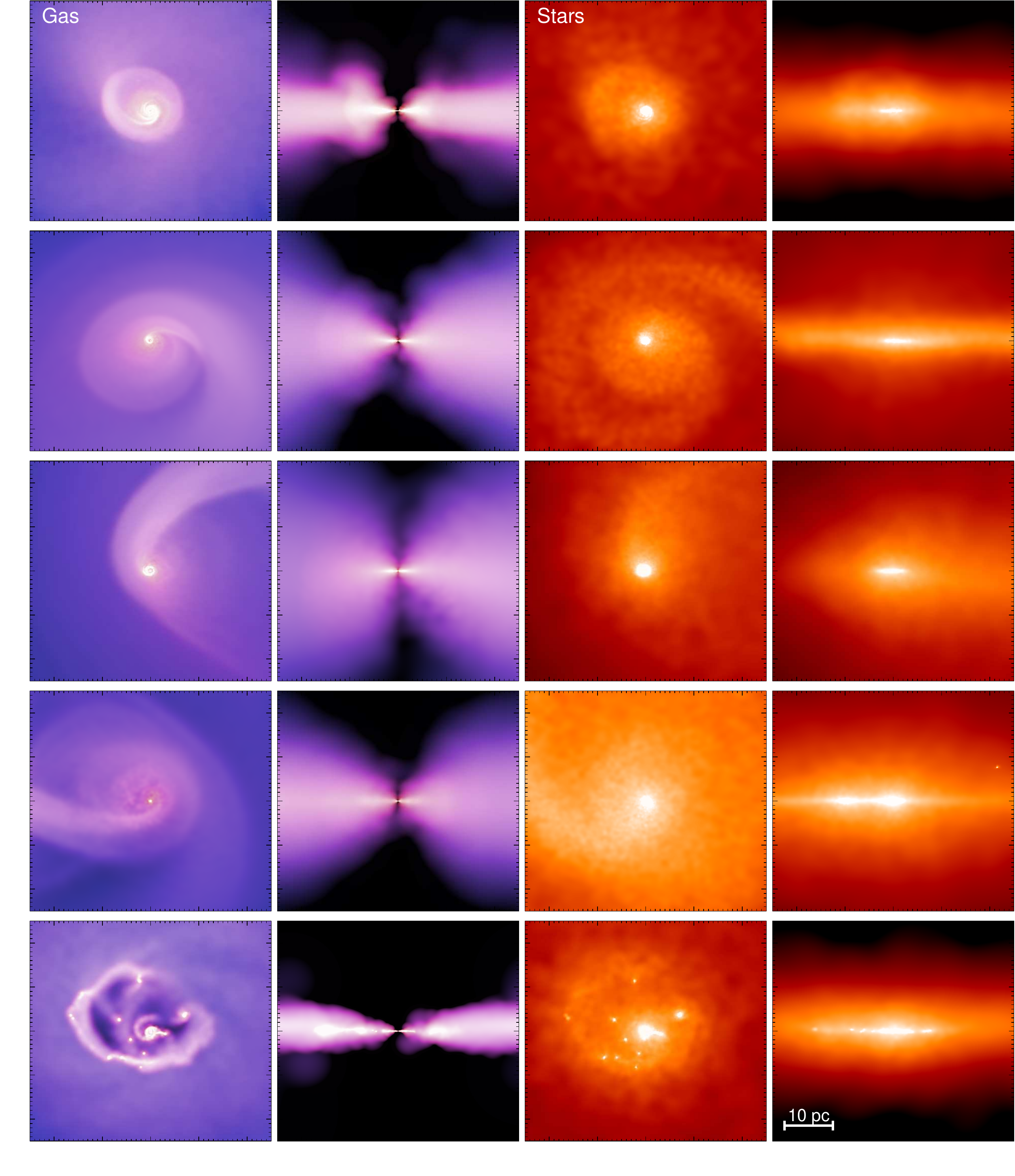}
    \caption{The nuclear disk in several representative simulations.
      Scale in all panels is the same (lower right). The different
      rows show the results from  simulations of the central
      $\sim 50$ pc of galactic nuclei  with
      different galaxy properties (top to bottom: Nf8h1c0thin,
      Nf8h1c1thin, Nf8h1c1qs, Nf8h1c1dens, Nf8h1c0 in
      \citealt{hopkins:zoom.sims}, which have initial $f_{\rm
        gas}\sim0.5-0.8$, $h/R=0.16,0.08,0.28,0.25,0.15$, $M_{\rm
        BH}\sim3\times10^{7}\,\msun$, and initial disky mass $\sim
      1.2,1.7,3.0,8.1,0.25\times10^{7}\,\msun$ inside $10\,$pc), and
      different treatments of stellar feedback (sub-grid sound speeds
      $c_{s}\sim35,\,20,\,40,\,50,\,10\,{\rm km\,s^{-1}}$ from top to
      bottom).  The formation of a lopsided disk is ubiquitous.  {\em
        Left:} Gas surface density.  Colors encode the absolute star
      formation rate of the gas (increasing from blue to red/yellow).
      Regions where gas shocks (edges in this image) dissipate energy,
      leading to rapid gas inflow.  {\em Middle Left:} Same, viewed
      edge-on in cylindrical ($R,\,z$) coordinates to emphasize the
      disk thickness versus radius.  The exact thickness depends on
      our model for stellar feedback, but gravitationally driven
      turbulence and heating results in the disks always being
      somewhat flared and thick on these scales, even with negligible
      stellar feedback ({\em bottom row}).  {\em Middle Right:} As
      {\em left}, but showing the stellar mass distribution.  The lack
      of shocks means that the edges of the disk are less sharp, but
      they are still visually clear.  {\em Right:} Edge-on stellar
      density ($x,\,z$).  Several of the systems appear to have
      two nuclei, which was the initial indication of the eccentric
      disk in M31 (the P1/P2 feature). 
      \label{fig:tile.10pc}}
\end{figure}

We have run a suite of over $\sim50$ simulations in order to study the
properties of, and robustness of, the eccentric disks in our
calculations. Figure~\ref{fig:tile.10pc} shows face-on and edge-on
views of the gas and stars in several of these simulations. Provided
that a significant amount of gas ($\gtrsim 10 \% \, M_{\rm BH}$) can
be driven in from larger radii by global torques in galaxies,
eccentric nuclear disks are generic.

\begin{figure}
    \centering
    \plotone{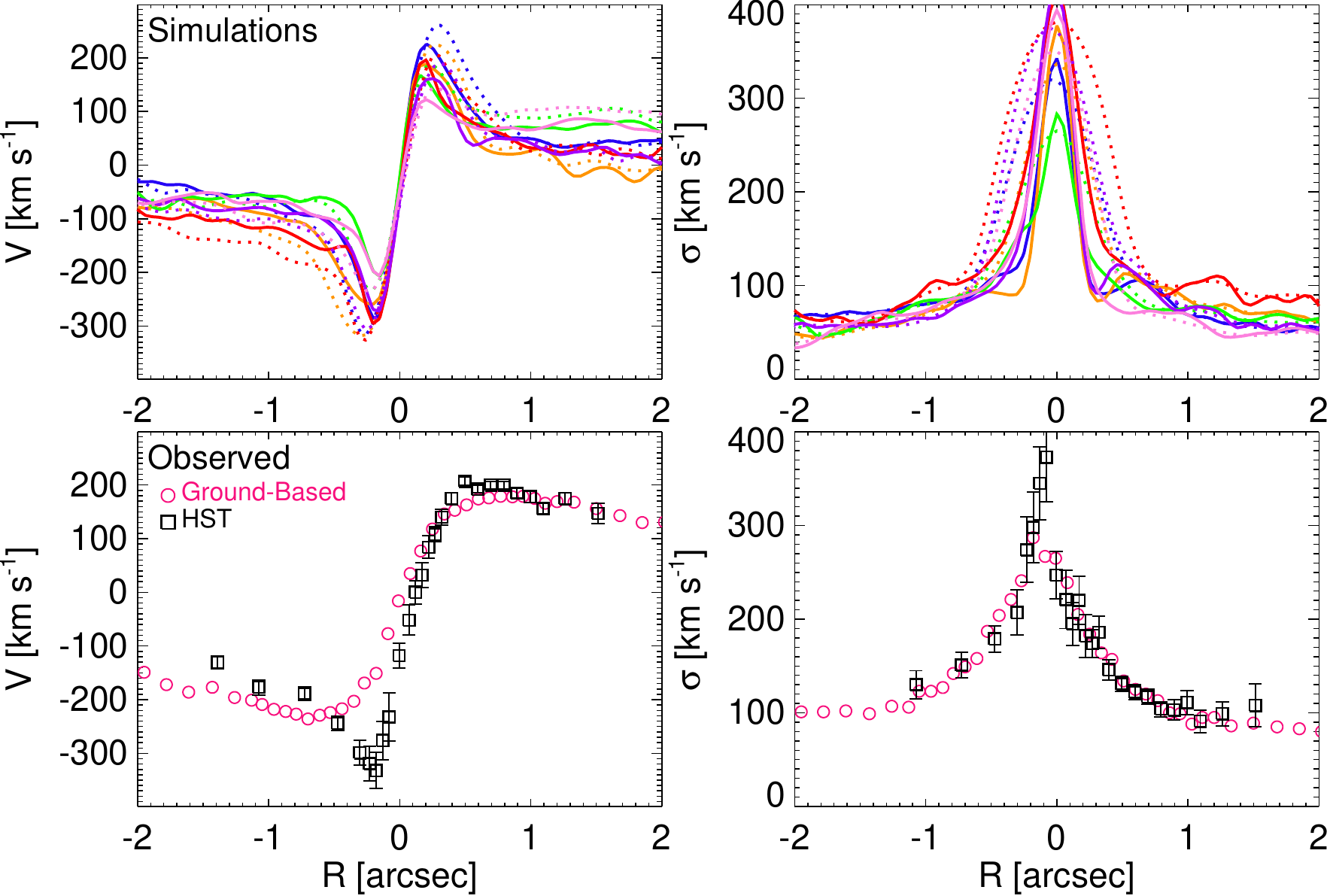}
    \caption{{\em Top:} Mean projected line-of-sight velocity $V$
      ({\em left}) and velocity dispersion $\sigma$ ({\em right}) of
      the relic nuclear stellar disks in our simulations, long after
      gas is exhausted.  Each solid line is a mock line-of-sight
      velocity field of the stars in one of our simulations, as if
      they were at the distance and viewing angle of the M31 nucleus
      ($54\,\deg$ from edge on;
      \citealt{peiris:m31.nuclear.disk.models}).  Dotted lines vary
      the inclination angle by $10\,\deg$.  The velocities are
      measured in narrow radial pixels along a slit placed along the
      major axis.  The slit is chosen to match exactly the pixel size,
      slit width, and resolution limits of the best current
      observations.  {\em Bottom:} The observations of M31. Our mock
      profiles are matched to the resolution of the {\em Hubble Space
        Telescope} observations shown here (black squares)
      \citep{bender:m31.nuclear.disk.obs}. The magenta circles are
      ground-based observations \citep{kormendybender:m31} which
      extend to larger radii but have inferior resolution and smooth
      the velocity field at $|R|<1\,$arcsec.
      \label{fig:velocities}}
\end{figure}

As the flow of gas through the nuclear region subsides, a stellar
remnant will remain behind that can retain the eccentric pattern.  The
characteristic radii $\sim 1-10$ pc and stellar masses $\sim 0.1-1 \,
M_{\rm BH}$ of the eccentric nuclear disks in our simulations are
reasonably consistent with those observed in M31 and other systems.
Figure~\ref{fig:tile.10pc} shows stellar density maps for several of
our simulations, including those that have exhausted most of their
gas.  The distinct nuclear disk is evident.  In an edge-on projection,
several of these nuclear disks appear to have double nuclei (secondary
brightness peaks) -- this is caused by the high density of stars near
apocenter in their elliptical orbits.  In several cases, this closely
resembles the secondary brightness peak in M31 (the P1 feature),
believed to arise in the same manner; note in particular the
second-from-bottom edge-on panel, where the two peaks are close in
brightness and separated by $\sim4$\,pc.  In our simulations we
confirm what has been inferred from dynamical models of P1, that the
appearance of such secondary nuclei is sensitive to projection
effects, and usually requires a sightline reasonably close to edge-on.

Figure~\ref{fig:velocities} shows the velocity field of the stars in
these relics, long after the gas is exhausted, and scaled as if
observed at the center of M31; the data for M31 are in the bottom
panel.  The overall agreement is impressive, particularly given that
these simulations are not designed to reproduce the observed features
of M31 in any way, but rather to study the growth of massive BHs.  We
find similar agreement when comparing to observations of the nuclear
disk in NGC 4486b \citep{lauer:ngc4486b}. The rotational velocity field of 
M31 is slightly less symmetric than our ``average'' simulation, but 
a number agree quite closely, and the level of dispersion asymmetry observed is typical for our simulations.


\begin{figure}
    \centering
    \plotone{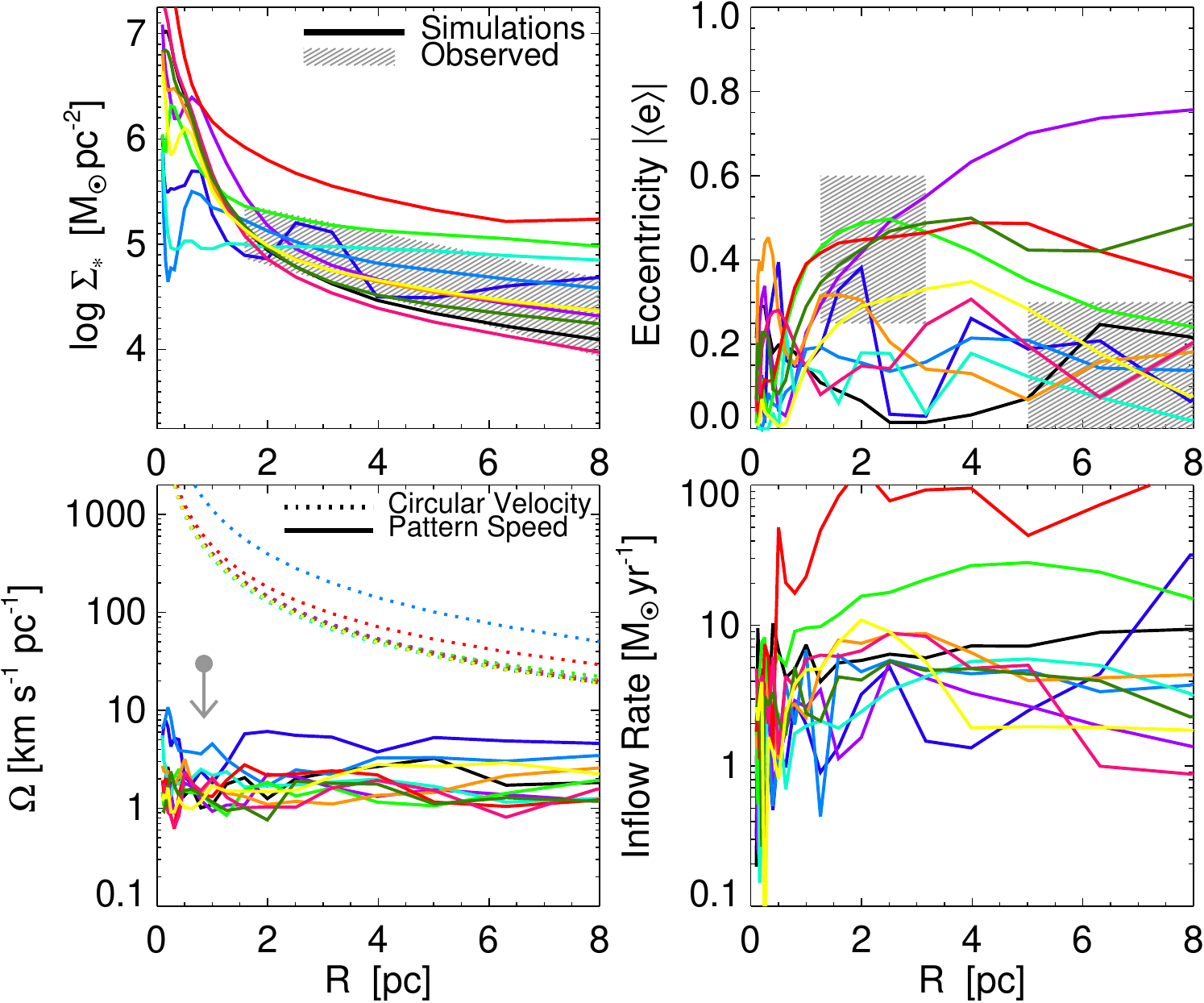}
    \caption{Properties of the simulated nuclear stellar disks
    versus radii, over the observed scales in M31. Each solid line corresponds 
    to a simulation, as in Figure~\ref{fig:velocities}.  
    Where available, we compare with the properties of the M31 
    system inferred from observations (grey shaded regions). 
    {\em Top Left:} Stellar mass surface density. 
    {\em Top Right:} Mean eccentricity of the disk along its major axis. 
    {\em Bottom Left:} Angular pattern speed (precession rate) 
    of the disk, which is much less than the angular velocity of 
    individual stars in the disk (dotted lines). 
    {\em Bottom Right:} Inflow rate of gas driven by the 
    gravitational torques of  the eccentric disk itself, during the active phase when the 
    disk formed. This accretion can produce most of the BHs growth.
    \label{fig:parameters}}
\end{figure}

Figure~\ref{fig:parameters} compares a number of the properties of our
simulated relic stellar disks with those of the M31 system inferred
from detailed kinematic studies
\citep{salow:nuclear.disk.models,jacobs:longlived.lopsided.disk.modes,
  sambhus:m31.nuclear.disk.model,peiris:m31.nuclear.disk.models,
  salow:nuclear.disk.models.2}.  The range of simulations shown
includes both variations in initial conditions and the treatment of
stellar feedback (the sub-grid turbulent velocity).
The plotted surface densities are azimuthally averaged (this suppresses 
the double-peaked appearance of the 
simulations and M31, but is more robust to projection effects). 
Note that the pattern speed $\Omega_{p}$ in our simulations is quite
low $\approx1-5 \, {\rm km\,s^{-1}\,pc^{-1}}$ (lower left), much less
than the rotation rate of individual stars at small radii.  In our
simulations, the pattern speed is set at the large radii where the
eccentric mode begins.
The actual precession rate in M31 is not very well-constrained, but
most studies place an upper limit of $<30\,{\rm km\,s^{-1}}$
\citep{sambhus:m31.pattern.speed}, and several studies imply a value
close to our prediction \citep[see][]{bacon:m31.disk}.  The mean disk
eccentricity in the simulations is also in broad agreement with that
observed, although we find that this is a less robust property of the
simulations and varies significantly from one simulation to another.

Figure~\ref{fig:parameters} also shows the inflow rates generated by
the nuclear stellar disk during its active/gas-rich phase, as a
function of radius. Nuclear disks that are similar to M31 in their
relic properties generate accretion rates up to several $\msun\,{\rm
  yr^{-1}}$ during the quasar epoch, when they are gas-rich.  This
highlights the key role that eccentric stellar disks can play in
fueling the growth of their host BHs.



Our models include a very simplified treatment of the feedback from
supernovae and massive stars: we introduce a sub-grid non-thermal
sound speed that is a proxy for the the effective turbulent speed of
the interstellar gas. To test the impact of this on our results, we
carried out calculations with identical initial conditions and
turbulent velocities ranging from $\sim10-100\,{\rm km\,s^{-1}}$,
roughly the lower and upper limits allowed by observational
constraints for the systems of interest (see Fig. 1 of
\citealt{hopkins:zoom.sims} and references therein).  The value of the
sub-grid sound speed has a significant effect on the amount of
resolved sub-structure in the simulation, with more sub-structure
present in simulations with lower turbulent velocities.  This is not
surprising since larger turbulent velocities raise the Jeans
mass/length, above which gravity is the dominant force.  However, all
of the simulations show a similar nuclear lopsided disk.  In terms of
the properties shown in Figures~\ref{fig:velocities} \&
\ref{fig:parameters}, the differences produced by changing the
sub-grid model are similar to the differences produced by somewhat
different galaxy properties. The fundamental reason for the weak
dependence on the subgrid model is that the torques in our simulations
are primarily determined by gravity, not hydrodynamic forces or
viscosity (see \citealt{hopkins:zoom.sims} for a detailed discussion).
The primary role of the subgrid feedback model is simply to prevent
catastrophic fragmentation of the galactic gas.



\vspace{-0.7cm}
\section{Discussion}
\label{sec:discussion}

\citet{hopkins:zoom.sims} argue that the dominant mechanism of angular
momentum transport in gas-rich galactic nuclei, from near the BH
radius influence ($\sim 10$ pc) down to the Keplerian viscous
accretion disk ($\ll 0.1$ pc), is gravitational torques produced by an
eccentric, lopsided disk (an $m = 1$ mode); this asymmetric disk forms
in our simulations when the disk mass is at least $\sim 10 \%$ of the
BH mass.
These torques provide the ``missing link'' connecting the gas
reservoir on galactic scales to the small-scale accretion disk near
the central BH.  In this {\em Letter} we have shown that the
long-lived (``fossil'') stellar relics of these disks are remarkably
similar to the eccentric stellar disk observed around the BH in M31.
This suggests that the stellar kinematics and morphology in galactic
nuclei can provide new insights into the physics of BH growth.

Emboldened by our success, we can use the observed properties of the
M31 disk to infer the accretion it was responsible for.  If the M31
disk was at one point gas-rich, the eccentric pattern in the stars
would produce strong torques in the gas, leading to an accretion rate
of $\dot M \sim \Sigma_{\rm gas}\,R^{2}\,\Omega\,|\Phi_{1}/\Phi_{0}|$
where $\Phi_{1}$ and $\Phi_{0}$ are the asymmetric and axisymmetric
terms in the gravitational potential, respectively.  For the measured
BH mass \citep[$10^{8}\,\msun$,][]{bender:m31.nuclear.disk.obs} and
potential of the eccentric disk \citep{peiris:m31.nuclear.disk.models}
this implies an accretion rate of $\sim 1 \, \msun\,{\rm yr^{-1}}$,
close to the Eddington limit of $2.4\,\msun\,{\rm yr^{-1}}$.  More
directly, we find that simulations that yield stellar relics in
closest agreement with M31 have typical inflow rates at $R\lesssim
0.1\,$pc in their active phases of $\sim0.3-5\,\msun\,{\rm yr^{-1}}$
(Fig. \ref{fig:parameters}).
These accretion rates imply that the observed stellar disk could have
helped the M31 BH gain much of its mass.

During the gas-rich phase, the typical column density of gas for an
edge-on line of sight through the disk is $N_{H} \sim 10^{25-26}\,{\rm
  atoms\,cm^{-2}}$, sufficient to obscure the radiation from the BH
even in the X-rays.  A combination of our model of stellar feedback
and self-consistently calculated gravitational perturbations generate
large `random' motions in the gas: the disks are thus thick, with
column densities sufficient to block the optical light ($N_{H} \gtrsim
10^{22}\,{\rm atoms \, cm^{-2}}$) out to an angle $\sim 20-45\deg$
above the plane -- in other words, a fraction $\sim30-60\%$ of all
sightlines will be obscured by gas and dust in the nuclear disk.  More
detailed conclusions about this obscuration will require a better
understanding of the role of gravitational heating and stellar
feedback in this unusual region.  Nonetheless, the properties of the
obscuring disk we infer are strikingly similar to those invoked for
the canonical ``toroidal obscuring region,'' assumed to reside on
small scales and to account for most of the optically obscured AGN
population \citep{antonucci:agn.unification.review,
  urry:radio.unification.review,lawrence:receding.torus}.  In the
context of our model, the observed ubiquity of the torus suddenly has
a dynamical origin: it itself helps {drive} the accretion.

Understanding the longevity of the eccentric disk in M31 has been as
challenging as understanding its origin.  The precession rate of the
disk is slightly different at different radii -- this should lead to
phase-mixing that ultimately wipes out the coherent eccentricity of
the disk. Indeed, we do see that the eccentric pattern damps away at
larger radii; however, the pattern at radii $\sim$pc, where the M31
disk is observed, persists in our simulations as long as they can be
reliably evolved, for $10^{8}$\,yrs, which is $\sim 10^{4}$ dynamical
times.  Self-gravity is likely to help maintain the pattern even
longer, in principle for much longer than the age of the universe
\citep{bacon:m31.disk,jacobs:longlived.lopsided.disk.modes,salow:nuclear.disk.models.2}.

Our simulations demonstrate that eccentric stellar and gaseous disks
form whenever the mass in the disky component in the central $\sim
10-30$ pc is comparable to that of the BH \citep{hopkins:zoom.sims}.
It is unclear, however, under what conditions these nuclear stellar
disks will survive to the present day.  The long-term stability of
isolated eccentric disks is not fully understood. In addition, it is
easy to imagine that subsequent ``dry'' galaxy-galaxy mergers might
destroy these features.  This is a key question for future research.
Observationally, there are a number of candidate systems in addition
to M31, as evidenced by apparently offset centers, ``hollow'' central
light profiles, double nuclei, or chemically distinct secondary nuclei
\citep{lauer:central.minimum.ell,lauer:centers,
  debattista:vcc128.binary.nucleus,thatte:m83.double.nucleus,
  afanasiev:2002.ngc5055.nuclear.disk}; and NGC4486b is another
confirmed eccentric disk \citep{lauer:ngc4486b}.  There may be similar
features in nuclear gas disks as well
\citep{seth:ngc404.nuclear.disk}.  But eccentric disks clearly do not
exist in all systems, as evidenced by the nucleus of M32.
More detailed observational constraints on the fraction of galaxies
with old, asymmetric nuclear stellar disks would provide a strong
constraint on our models.

When nuclear eccentric disks do survive, our models imply that there
should be a correspondence between the properties of the nuclear
stellar disk (mass, radius, and asymmetry) and the central BH mass, as
we have demonstrated is the case in M31. Very low pattern speeds
$<10\,{\rm km\,s^{-1}\,pc^{-1}}$ should be ubiquitous, as they are
required for efficient exchange of angular momentum between the stars
and gas.  It is also worth noting that if some gas flows in from
larger radii, or accumulates via stellar evolution, the nuclear
regions can experience recurrent low-level AGN activity and/or star
formation, regulated by the same mechanism of eccentric stellar torques
(e.g., M31's young stellar population;
\citealt{chang:m31.eccentric.disk.model}).
Such episodes may complicate dating the formation epochs of these
disks, but also provide a laboratory to study the physics of inflow in
detail.  The properties of the nuclear disk in its gas-rich phase
determines the distribution of implied torus scale lengths and gas
densities, which can be probed by infrared adaptive optics
observations of nearby bright AGN
\citep{davies:sfr.properties.in.torus,hicks:obs.torus.properties}.  On
the theoretical side, further improvements in the treatment of gas
physics and star formation will enable more detailed comparison with
observations.

\vspace{-0.5cm}
\acknowledgments We thank Phil Chang, Lars Hernquist, Scott Tremaine,
John Kormendy, and Tod Lauer for helpful discussions during the
development of this work. Support for PFH and EQ was provided by the
Miller Institute for Basic Research in Science, University of
California Berkeley. EQ was also supported in part by NASA grant
NNG06GI68G and the David and Lucile Packard Foundation.
\\

\bibliography{/Users/phopkins/Documents/lars_galaxies/papers/ms}

\end{document}